\documentclass[aps,prl,twocolumn,groupedaddress,superscriptaddress,showpacs,floatfix]{revtex4-1}
\usepackage{graphicx}
\usepackage{xcolor}
\usepackage{amsfonts}
\usepackage{amsmath}
\usepackage{amssymb}
\usepackage{rotating}
\usepackage[english]{babel}
\usepackage{physics}
\providecommand{\abs}[1]{\lvert#1\rvert}

\begin{document}

\title{Unconventional spin texture in a non-centrosymmetric quantum spin Hall insulator} 
\author{C. Mera Acosta}
\email{acosta@if.usp.br}
\affiliation{Instituto de F\'isica, Universidade de S\~ao Paulo, CP 66318, 05315-970, S\~ao Paulo, SP, Brazil}
\author{O. Babilonia} 
\affiliation{Instituto de F\'isica, Universidade de S\~ao Paulo, CP 66318, 05315-970, S\~ao Paulo, SP, Brazil}
\author{L. Abdalla} 
\affiliation{University of Colorado, Boulder, Colorado 80309, USA}
\author{A. Fazzio} 
\affiliation{Instituto de F\'isica, Universidade de S\~ao Paulo, CP 66318, 05315-970, S\~ao Paulo, SP, Brazil}
\affiliation{Centro de Ci\^encias Naturais e Humanas, Universidade Federal do ABC, Santo Andr\'e, CP 09210-170, S\~ao Paulo, Brazil.}
\date{\today}

\begin{abstract}
We proposed that the simultaneous presence of both Rashba and band inversion can lead to a Rashba-like spin-splitting formed by two bands with the same in-plane helical spin texture. Because of this unconventional spin texture, the backscattering is forbidden in edge and bulk conductivity channels. We propose a new non-centrosymmetric honeycomb-lattice quantum spin Hall (QSH) insulator family formed by the IV, V, and VII elements with this property. The system formed by Bi, Pb and I atoms is mechanically stable and has both a large Rashba spin-splitting of 60 meV and a large nontrivial band gap of 0.14 eV. 
Since the edge and the bulk states are protected by the TR symmetry, contrary to what happens in most doped QSH insulators, the bulk states do not contribute to the backscattering in the electronic transport, allowing the construction of a spintronic device with less energy loss.
\end{abstract}

\maketitle

The main objective of spintronics is to understand the mechanisms by which it is possible to achieve efficient control of both spin configurations and spin currents\cite{Dmytro2012,DasSarma2004}. In the last decade, the way to achieve this objective has experienced a breakthrough due to \textit{i}) the discovery and understanding of mechanisms to generate spin currents in conductors with magnetic order and in paramagnetic conductors/semiconductors\cite{Arne2012,Tian2015,Luqiao2012}, \textit{ii}) the experimental observation of theoretically proposed spin injector systems\cite{Hasan2012,Qi2011,Brune2012}, and \textit{iii}) the synthesis of 2D materials with long spin relaxation time\cite{Dmytro2012,Kulik2015}. The generation of spin currents, spin injections and spin conservation are mediated by the spin-orbit coupling (SOC) mainly via Rashba effect and/or nontrivial topological phases\cite{SpinCurrent2012,Ron2012,Manchon2015,Bercioux2015}, such as the quantum spin Hall (QSH) effect\cite{Kane2005}. Therefore, the search for systems experiencing these properties is a primary concern for the development of spintronics.

QSH insulators support helical metallic edge states, forming topological Dirac fermions protected by the time-reversal (TR) symmetry on an insulating bulk\cite{Hasan2012, Qi2011}. The topological transition from trivial insulating to topological insulators is evidenced as a band inversion at the TR invariant $k$-point mediated by the SOC.
The topological band dispersion has been experimentally characterized via angle-resolved photoemission spectroscopy (ARPES) and local scanning tunneling microscopy (STM) in 3D topological insulators\cite{Hasan2012}, and via transport measurements in HgTe/CdTe quantum wells\cite{Konig02112007,Olshanetsky2015}. On the other hand, the Rashba effect, arising from the lack of inversion symmetry, leads to parallel spin-polarized band dispersion curves with opposite in-plane chiral spin texture\cite{Rashba1984}, allowing the control of the spin direction through an electric field\cite{SpinCurrent2012,Manchon2015,Bercioux2015}.
These dispersion curves and Fermi contours have been characterized by spectroscopic measurements for many surfaces and interfaces\cite{LaShell1996,Ast2007,Koroteev2004,Nitta1997}. Large Rashba spin-splitting are found in materials formed by heavy elements with strong intrinsic SOC such as Bi, Pb, W, among others\cite{Hirahara2006a,Mathias2010,Nitta1997,Hongtao2013,Dil2008}. In this work, we look at the consequences of the simultaneous presence of a Rashba spin-splitting and a inverted bandgap. Such properties appear simultaneously in thin films and heterostructures of 3D topological insulators\cite{Zhu2011,ZhangYi2010,Ishizaka2011,Bahramy2012,Das2013,Zhou2014}.

Here, we show that bulk states can be protected against backscattering in nanoribbons of QSH insulators with bulk inversion asymmetry. This behavior is a consequence of the simultaneous presence of both Rashba and band inversion in a QSH insulator. In our model, both the conduction and the valence bands are formed by two bands with the same in-plane helical spin texture and opposite $\langle S_{z}\rangle$ spin component.
We propose a stable non-centrosymmetric honeycomb-lattice QSH insulator that presents this unconventional bulk spin texture. This system is formed by the Bi, Pb, and I elements and, has a large nontrivial band gap of $0.14$ eV and a huge Rashba spin-splitting of $60$ meV. To construct the Hamiltonian exhibiting the proposed spin texture we will use the PbBiI system.
\begin{figure}
\includegraphics[width = 8.6cm]{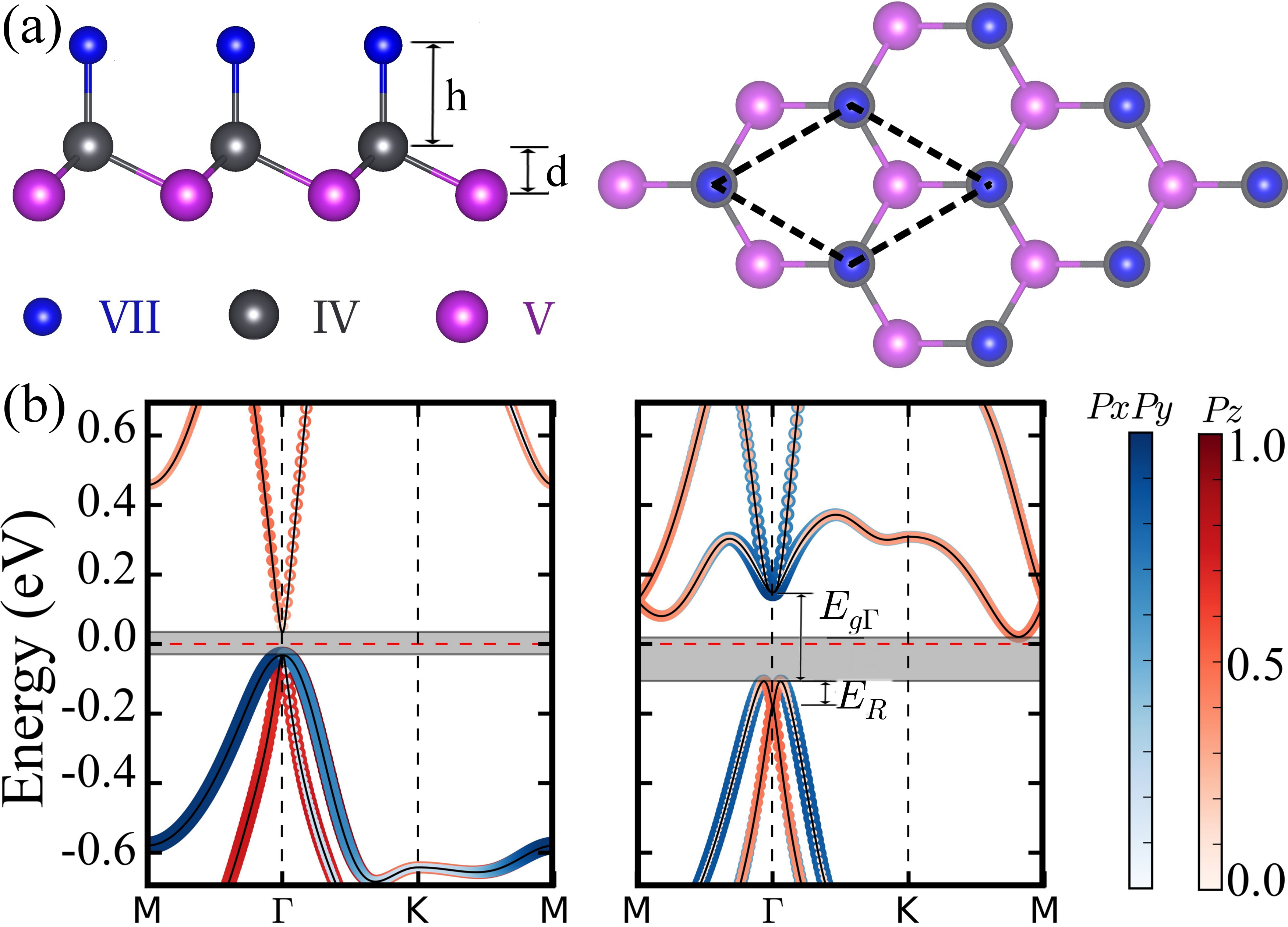}
\caption{(color online). (a) Top and side view of the PbBiI atomic structure.  In the lowest energy configuration the buckled, $d$, has 1.3 \AA\ in length and, the Bi-Pb and Pb-I ($h$) bounds have 3.04 \AA\ and 1.35 \AA\ in length, respectively. (b) Band structure without SOC (left) and with SOC (right). The color scales represent the weight of the orbital projection in the wavefunction  $\psi_{n}(\vec{k})$. The  projections in the $p_z$($p_x$ and $p_y$) Bi orbitals are indicated by red(blue). The Rashba spin-splitting and the band inversion are characterized by $E_{R}$ and $E_{g \Gamma}$, respectively.}
\label{Fig1}
\end{figure}

Figure~\ref{Fig1} summarizes the crystalline structure and the results we obtain from \textit{ab initio} calculations, which are performed within the density functional theory (DFT) framework as implemented in the SIESTA code\cite{soler2002siesta} and in the Vienna Ab Initio Simulation Package\cite{PhysRevB.54.11169}. We consider the on-site approximation for the SOC\cite{PhysRevB.89.155438,fernandez2006site} in the SIESTA code. The Local Density Approximation\cite{perdew1981self} and the Perdew-Burke-Ernzenhof generalized gradient approximation\cite{PhysRevLett.77.3865} are used for the exchange-correlation functional.
Interpreting the hexagonal lattice as two triangular sub-lattices A and B,  the system has a V atom type on the sub-lattice A, and a IV-VII dimer in the sub-lattice B (Fig \ref{Fig1}a). These non-centrosymmetry systems have a buckled format and fulfill the symmetry operations of the $C_{3v}$ symmetry: \textit{i}) three-fold rotation symmetry $R_3$ along the $z$ axis, \textit{ii}) mirror symmetry $M_{x}$ ($x\to -x$) in the $yz$ plane, and \textit{iv}) TR symmetry $\mathcal{T}$. We predict that the PbBiI system is mechanically stable, imaginary frequencies does not exist in the \textit{ab initio} phonon spectrum (see Supplemental Material) and the formation energy ($E_{F}=E_{\textrm{PbBiI}}-\mu_\textrm{Pb}-\mu_\textrm{Bi}-\mu_\textrm{I}$) is about $-0.77$ eV.  

At the $\Gamma$ point, the top of the valence band is dominated by the $p_{x,y}$ Bi orbitals and the bottom of the conduction band mainly consists of the $p_{z}$ Bi orbitals, as shown in Fig \ref{Fig1}b. 
When the SOC is taken into account, the $p$ orbitals are mixed to form effective orbitals preserving the total angular momentum and a band inversion occurs when $\lambda_{\textrm{SOC}}=0.65$, where $\lambda_{\textrm{SOC}}=0(1)$ means the absence (full presence) of SOC.
We implemented the evolution of Wannier center of charges as an alternative method to the $Z_{2}$ invariant calculation using \textit{ab-inito} simulations \cite{Yu2011,PhysRevB.83.035108,PhysRevB.83.235401}. We find that there is no a horizontal reference line that crosses the evolution of the WCCs at least an odd number of times (see Supplemental Material), showing a value of $Z_{2}=1$\cite{Yu2011,PhysRevB.83.235401}, and hence, confirming that the PbBiI system is a QSH insulator.
On the other hand, according to the symmetry operation, the wavefunction at the $\Gamma$ point is given by the $\{|\Lambda_{J},j_{z}\rangle\}$ effective states, where $J$ is the total angular momentum, $j_{z}$ is the projection along the $z$ axes, and $\Lambda$ corresponds to the Bi and Pb-I contributions. To preserve the total angular momentum, the $|\Lambda_{3/2},\pm 3/2\rangle$ effective states should be a linear combination of  the $p_{+}=p_{x}+ip_{y}$ and $p_{-}=p_{x}-ip_{y}$ effective orbitals and, the $|\Lambda_{J},\pm 1/2\rangle$ effective states should be a linear combination of the $p$ orbitals, mainly $p_z$ orbitals.  In this inverted band gap the conduction band mainly consists of $p_{x,y}$ Bi orbitals and the valence band is formed by the $p$ orbitals, mainly dominated by $p_{z}$ Bi orbitals, as shown in Fig \ref{Fig1}b. Therefore, at the $\Gamma$ point, the valence (conduction) band is described by the effective states $\{|\textrm{Bi}_{J},j_{z}\rangle\}$ with $J=3/2$ ($J=1/2$)  and hence, we write the Hamiltonian using the full SOC basis $\{|\textrm{Bi}_{1/2},1/2\rangle, |\textrm{Bi}_{1/2},-1/2\rangle, |\textrm{Bi}_{3/2},1/2\rangle, |\textrm{Bi}_{3/2},-1/2\rangle\}$. 

\begin{figure}
\begin{center}
\includegraphics[width = 8.6cm]{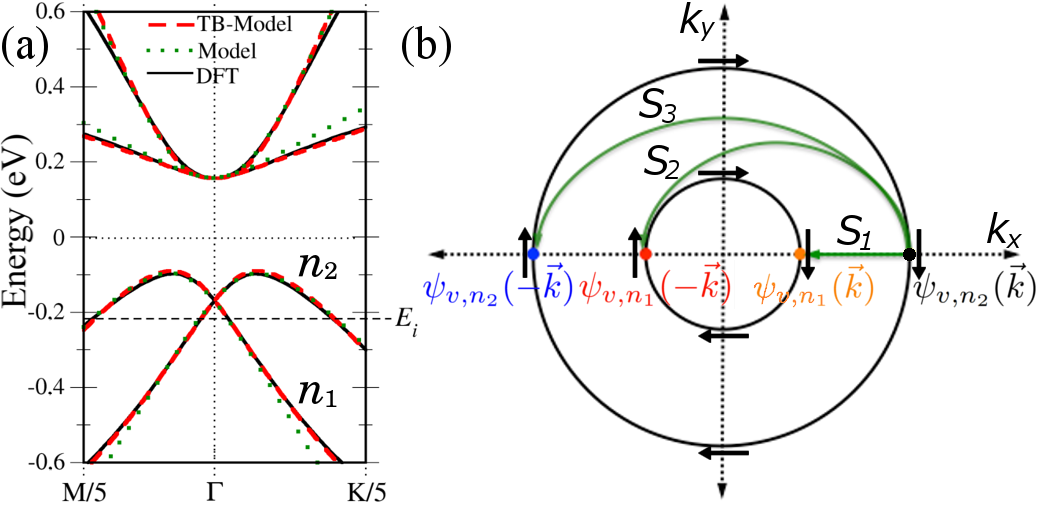}
\caption{(color online). (a) Band structure calculated with DFT, complete tight-binding model and simplified model. (b) Fermi contour at the energy plane $E_{i}$ obtained from the simplified model. The backscattering processes and the spin textures are represented by green and black arrows, respectively.}
\label{FigScatt}
\end{center}
\end{figure}

The tight-binding Hamiltonian matrix elements 
are given by:
\begin{equation}
[\mathcal{H}(\vec{k})]_{ij}=\varepsilon_{ij}\delta_{ij}+\sum_{\nu=1}^{6}t^{ij}_{\vec{a}_{\nu}}e^{i\vec{k}\cdot\vec{a}_{\nu}},
\label{matrixelements}
\end{equation}
where $i=(\textrm{Bi},J,j_{z})$, $j=(\textrm{Bi},J',j_{z}')$ and $\varepsilon$ is the on-site energy. Since the $|\textrm{(Pb-I)}_{J},j_{z}\rangle$ effective states contribution is not relevant near the Fermi energy, we omit the terms associated with the nearest neighbors (Pb-I dimer) and hence, $t^{ij}_{\vec{a}_{\nu}}=\langle\vec{n},\textrm{Bi}_{J},j_{z}|H|\vec{a}_{\nu},\textrm{Bi}_{J'},j_{z}'\rangle$ represents the next nearest neighbor hopping terms, with $\vec{n}$ indicating the lattice site and $\vec{a}_{\nu}$ corresponding to the $\nu$-th of the six next nearest neighbor vectors.  Different form buckled honeycomb lattice systems, such as Germanene, Silicene among others, in which the nearest neighbor hopping terms are essential to its description, in the PbBiI the Pb-I dimer only mediates the interaction between Bi atoms and its effect is effectively introduced within the next nearest neighbor hopping terms. Therefore, the PbBiI Hamiltonian is striking different from the Kane-Mele model\cite{Kane2005}. Using the relevant symmetry operations of the $C_{3v}$ point group, these hopping terms can be related to each other and are uniquely determined (see Supplemental Material), which leads to an approximate description of the DFT band structure (see Fig \ref{FigScatt}a).

Considering the $\vec{k}\rightarrow\Gamma$ limit we obtain a reduced form for the tight-binding Hamiltonian matrix elements, 
\begin{equation}
 \mathcal{H}_{J}(\vec{k})=(-1)^{2J+1}\varepsilon_{J}+(-1)^{2J}h_{J,0}(\vec{k})+h_{J,R_{3}}(\vec{k})+h_{J,R_{1}}(\vec{k})
 \end{equation}
where $h^{J}_{0}(\vec{k})=\xi\vec{k}^{2}$, $h_{R_{3}}^{J}(\vec{k})=\alpha_{R_{3}}^{J}[(k_{+})^{3}+(k_{-})^{3}]\sigma_{z}$, $h_{R_{1}}^{J}(\vec{k})=\alpha_{R_{1}}^{J}(\vec{\sigma}\times\vec{k})\cdot\hat{z}$ and $\mathcal{H}_{int}=\gamma\vec{\sigma}\cdot \vec{k}$. Here, $\mathcal{H}_{1/2}(\vec{k})$ and $\mathcal{H}_{3/2}(\vec{k})$ are the effective terms that described the $|\textrm{Bi}_{1/2},\pm 1/2\rangle$ and $|\textrm{Bi}_{3/2},\pm 3/2\rangle$ states, respectively and $\mathcal{H}_{int}(\vec{k})$ is the interaction between these states.
The parameters 
are related to the hopping terms and are obtained via a least squares standard approach in order to match the DFT calculation (see Supplemental Material). 
Since we find that $\alpha_{R_{1}}^{3/2}(\vec{k})\approx 0$ and considering  $\xi^{1/2}\approx\xi^{3/2}=\xi$, we rewrite the Hamiltonian as 
\begin{equation}
\mathcal{H}(\vec{k})=\left(\begin{array}{cccc}
-\varepsilon+\xi k^{2}& i\alpha_{R_{1}}k_{-} & 0 & \gamma k_{-}\\
-i\alpha_{R_{1}}k_{+} & -\varepsilon+\xi k^{2}& \gamma k_{+} & 0\\
0 &  \gamma k_{-} & \varepsilon-\xi k^{2} & 0 \\
 \gamma k_{+} & 0 &0 & \varepsilon-\xi k^{2}
\end{array}\right).
\label{Halmiltonian}
\end{equation}
 
We plot the Fermi contours obtained from this Hamiltonian and represent the backscattering processes in Fig \ref{FigScatt}b. 
In the valence band, an energy plane below(above) the band crossing consists of two concentric circles with the same(opposite) in-plane helical spin texture. 
Likewise, in the conduction band, an energy plane consists of two concentric circles with the same in-plane helical spin texture. Because of this bulk spin texture, the elastic and inelastic backscattering processes represented by $S_{2}$ and $S_{3}$, respectively, 
are forbidden. 

\begin{figure}
\includegraphics[width = 8.6cm]{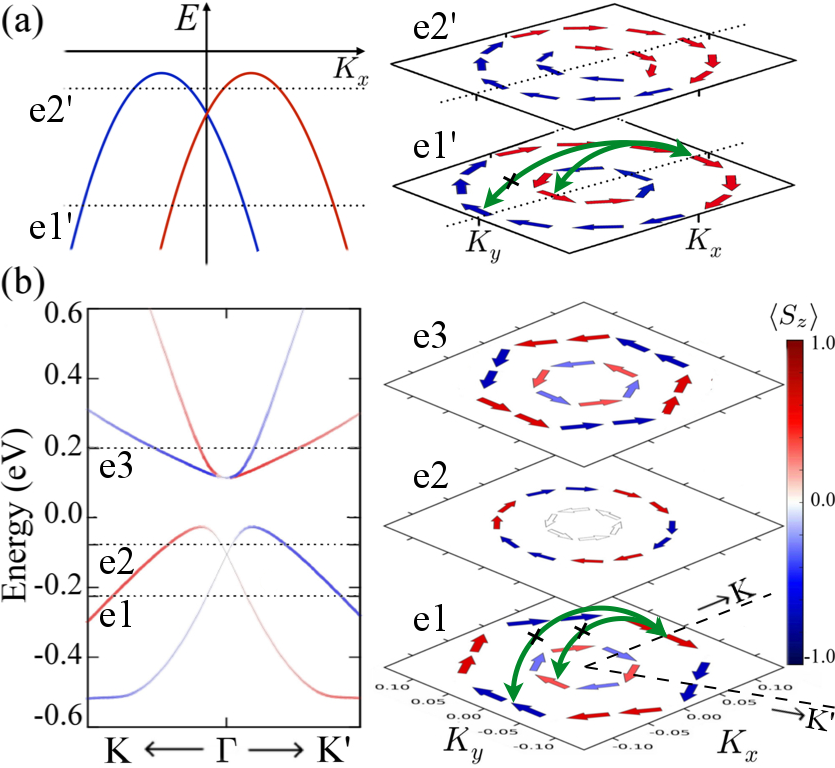}
\caption{(color online). Band structure and spin texture of (a) an out-plane spin polarized Rashba material and (b) the PbBiI system. 
The dotted lines in the band structure represent the energy planes (e1-e3 and e1'-e2') in which we show the in-plane spin texture, whose magnitudes are represented by the size of the arrows. In this arrows and in the band structure, the color code stands for the normalized $\langle S_{z}\rangle$ spin polarization. 
The backscattering processes $S_{2}$ and $S_{3}$ are represented by green arrows.
}
\label{Fig2}
\end{figure}

In order to quantify the probability of backscattering, we calculate the scattering rate due to a single coulomb impurity considering the bare coulomb potential\cite{Yin2014}, $\mathcal{S}_{\vec{k}'n',\vec{k}n}=\frac{2\pi}{\hbar}\frac{q^{4}_{e}}{4A^{2}\kappa^{2}\beta^{2}}(1-cos\theta_{\vec{k}',\vec{k}})\boldsymbol{I}_{\vec{k}',n',\vec{k},n}\delta(E_{n}-E_{n'})$. Here, $A$ is the unit area, $q_{e}$ is the single-electron charge, $\kappa$ is the static dielectric constant and $\beta=|\vec{k}-\vec{k}'|$ and $I_{\vec{k}',n_{1},\vec{k},n_{2}}=|\langle \psi_{E_{n_{1}}}(\vec{k}')|\psi_{E_{n_{2}}}(\vec{k})\rangle|^{2}$ is the overlap integral, which is calculated using the normalized wavefunction,
\begin{equation}
\psi_{E_{n_{\lambda}}}(\vec{k})=\sqrt{N}\left(\begin{array}{c}
1\\
-i\frac{k_{+}}{\alpha_{R_{1}}k^{2}}\frac{(\varepsilon-\xi k^{2})^{2}-E_{n_{\lambda}}^{2}+\gamma^{2}k^{2}}{\varepsilon-\xi k^{2}-E_{n_{\lambda}}}\\
i\frac{\gamma}{\alpha_{R_{1}}}\frac{(\varepsilon-\xi k^{2})^{2}-E_{n_{\lambda}}^{2}+\gamma^{2}k^{2}}{(\varepsilon-\xi k^{2}-E_{n_{\lambda}})^{2}}\\
\frac{-\gamma k_{+}}{(\varepsilon-\xi k^{2}-E_{n_{\lambda}})}\\
\end{array}\right),
\end{equation}
where $N=\frac{\abs{\varepsilon-\xi k^{2}-E_{n_{\lambda}}}^2}{2[(\varepsilon-\xi k^{2}-E_{n_{\lambda}})^{2}+\alpha_{int}k^{2}]}$ . We verify that $I_{-\vec{k}',n_{1},\vec{k},n_{2}}=I_{-\vec{k},n_{\lambda},\vec{k},n_{\lambda}}=0$ and therefore the scattering rates $\mathcal{S}_{-\vec{k}',n_{1},\vec{k},n_{2}}$  and $\mathcal{S}_{-\vec{k},n_{\lambda},\vec{k},n_{\lambda}}$ are null, proving that backscattering processes are unlikely. On the other hand, in an ordinary out-plane spin polarized Rashba material, such as the thin films of the BiTeI 3D topological insulator\cite{Ishizaka2011}, in an energy plane below(above) the band crossing the bands have opposite(same) in-plane chiral spin texture (Fig \ref{Fig2}a) and therefore, the elastic backscattering $S_{3}$ is forbidden. Different from PbBiI, in a Rashba semiconductor, below the band crossing the inelastic backscattering $S_{2}$ is allowed and $I_{-\vec{k}',n_{1},\vec{k},n_{2}}\approx 1$\cite{Rashba1984,Sakamoto2013}, as represented by the green arrows in Fig \ref{Fig2}a.

Far from the $\Gamma$ point, the $R_{3}$ symmetry breaking generates nonlinear terms in the SOC such as the three order Rashba term $h_{R_{3}}(\vec{k})$, which induces $\langle S_{z}\rangle$ spin polarization and hexagonal warping effect in the bulk states\cite{Vajna2012,Fu2009}, as represented in Fig \ref{Fig2}b. 
Although $S_{z}$ spin-polarized increases, the expected value of $S_{z}$ is still near zero. Because the $\langle S_x\rangle$ and $\langle S_y\rangle$ spin flips are required so that the inelastic scattering process $\mathcal{S}_{-\vec{k}',n_{1},\vec{k},n_{2}}$ occurs (see Fig \ref{Fig2}b), the overlap integral $I_{-\vec{k}',n_{1},\vec{k},n_{2}}$ is still small compared to the value found in an ordinary Rashba semiconductors and the probability that the scattering process $S_{2}$ occurs remains low.
Analogous to the surface states of Bi$_2$Te$_3$\cite{Fu2009}, near the $\Gamma$ point, the $\langle S_{z}\rangle$ spin-polarization and the warping hexagonal tend to zero and the only contribution coming from the lack of inversion symmetry is the first order Rashba term, $h_{R_{1}}(\vec{k})$. Thus, to obtain the unconventional spin texture near the $\Gamma$ point is only enough to consider both Rashba effect and band inversion, as it was done in eq. \ref{Halmiltonian}.
According to our DFT results, the Rashba spin-splitting is about $60$ meV, which are huge compared with the values found in semiconductors and surprisingly is among the highest found in 3D systems\cite{Hirahara2006a,Mathias2010,Nitta1997,Hongtao2013,Dil2008,Ishizaka2011}.
This value can be increased up to $E_{R}\approx 90$ meV applying large compressive strain (see Supplemental Material).

\begin{figure}
\includegraphics[width = 8.6cm]{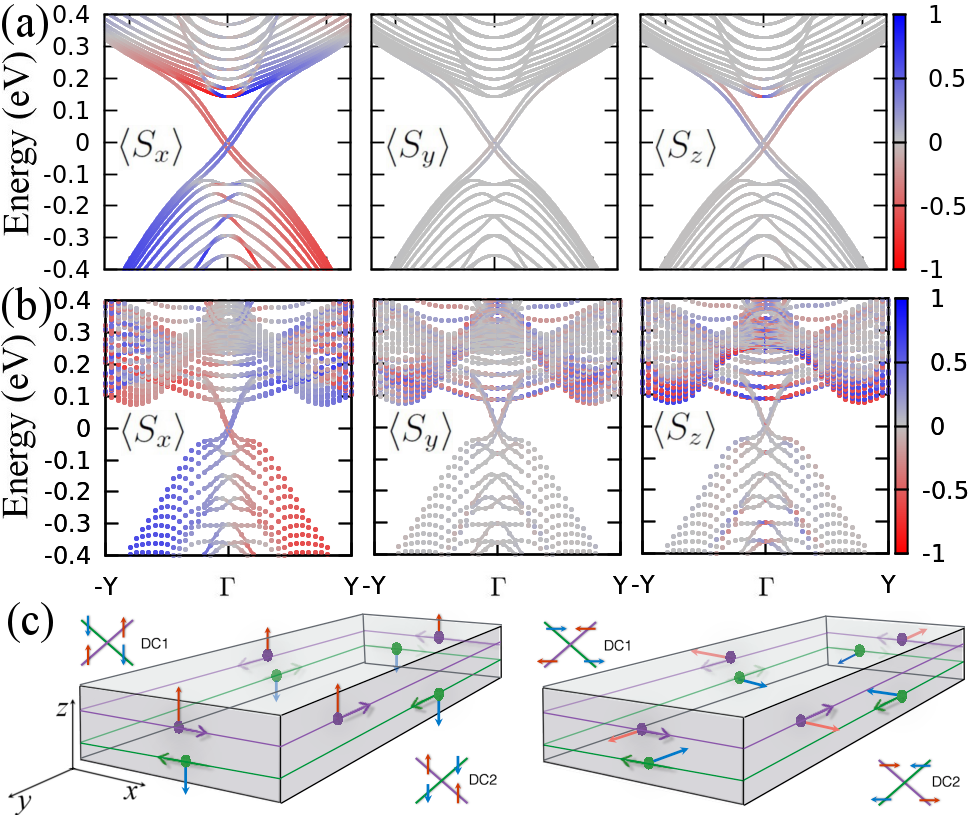}
\caption{(color online). (a) Tight-binding and (b) DFT spin-texture band structure of an armchair nanoribbon with 98.7 \AA~wide. The color code stands for the spin polarization. (c) Pictorial representation of edge states with $\langle S_{z}\rangle$ (left) and $\langle S_{x}\rangle$ (right) spin polarization.}
\label{Fig4}
\end{figure}

On the other hand, since the out-plane spin polarization oscillates according to the $C_{3v}$ symmetry, as occurs in thin films of Bi$_2$Te$_3$\cite{Fu2009}, at the $\Gamma\rightarrow M$ symmetry path, the $S_{z}$ spin component is zero (see Fig \ref{Fig2}) and therefore, inelastic backscattering processes are completely suppressed. The armchair nanoribbon BZ is parallel to the $\Gamma\rightarrow M$ symmetry path at the $k_{y}$ axis of the hexagonal BZ. Thus, scattering processes are dominated by the $S_{x}$ spin component and hence, the elastic and inelastic backscattering is forbidden for bulk and edge states, as shown in Fig \ref{Fig4}.
Similarly, the zigzag nanoribbon BZ is parallel to the $k_{x}$ axis and therefore, $\langle S_{x}\rangle =0$. Because of the non-zero $\langle S_{z}\rangle$ spin components, there is a low probability of inelastic backscattering in accordance with the bulk behavior discussed above (see Supplemental Material). Because of the strong SOC, the spin and momentum are constrained to be perpendicular. This spin-momentum locking implies that Dirac cones of different edges are required to have the same $S_{x}$ spin texture and different $S_{z}$ spin texture of spin in the armchair nanoribbon, as represented in Fig \ref{Fig4}c.

The protected bulk states near the $\Gamma$ point only appear in the energy region in which the Rashba effect and the band inversion point take place. This energy region overlaps with the bottom of the conduction band, which is at $\Gamma\rightarrow M$ symmetry path, as shown inf Fig \ref{Fig4}b. 
The bottom of the conduction band and hence, the energy range in which the unusual spin texture is present can be modified by applying tensile strain (see Supplemental Material). Indeed, the inverted bandgap at $\Gamma$ can be equal to the bandgap $E_{g}$ when strain is applied.

It is well established that the Bi-Pb alloy can be realized experimentally maintaining the $R\bar{3}m$ space group \cite{Gokcen1992,Huang1983}. The Pb-Bi rhombohedral alloy along the [111] direction can be considered as a stack of PbBi honeycomb lattices that are weakly bonded (mainly ruled by Van der Waals type interaction) to each other, similarly to the bismuth bilayers\cite{Drozdov2014}. The dangling bonds that appear at the Pb-rich PbBi surface can be eliminated by bonding to iodine atoms and hence, the proposed spin texture could be observed in the PbBiI system via STM experiments analogously to the observation of Bi-bilayers' edge states\cite{Drozdov2014}.

The $C_{3v}$ symmetry in the PbBiI system leads to a interaction term different from the BHZ model used to describe the QHS phase in HgTe/CdTe quantum wells\cite{Bernevig1757}. Although the BHZ model considering the Rashba effect has been used to describe asymmetric InAs/GaSb/AlSb quantum wells\cite{Liu2008}, the consequences of a huge Rashba spin-splitting and the three order Rashba term in a bulk inverted band gap, such as the unconventional spin texture reported here, have been ignored.

In summarizing, 
the simultaneous presence of a huge Rashba effect and a inverted bandgap in systems with $C_{3v}$ symmetry leads to conduction and valence bands with a Rashba-like spin-splitting with the same helical in-plane spin texture and with null $S_{z}$ spin-polarization at the $\Gamma\rightarrow M$ symmetry paths. Thus, the spin texture in the nanoribbons depends on its orientation. 
We find that bulk states are protected by the TR symmetry and contrary to what happens in most doped QSH systems, the bulk states do not contribute to the backscattering, opening the way for realizing novel applications of topological edge states.
Additionally, we proposed a new honeycomb-lattice QSH insulator mechanically stable - the PbBiI system, which has a large Rashba splitting of 60 meV, a large nontrivial gap of 0.14 eV and hence, it presents the predicted unconventional spin texture. 
As far as we know, the PbBiI system is the first system that has such spin texture properties in its bulk band structure.

\begin{acknowledgements}
This work was supported by the Sao Paulo research foundation (grant 2014/12357-3). We would like to thank Dr. Luis G. G. V. Dias da Silva by the discussions. We also acknowledge Dr. Soluyanov and Dr. Vanderbilt for sharing the code to calculate the WCCs with VASP.
\end{acknowledgements}

\bibliography{Ref}

\end{document}